\documentclass{aa}            
\usepackage{graphicx}
\begin{document}
\thesaurus{ }     
\title{The enigmatic outburst of V445 Puppis - a possible Helium nova?}
\author{ N. M. Ashok and D. P. K. Banerjee }
\offprints{N.M. Ashok}
\institute{Physical Research Laboratory, Navrangpura, \\
	      Ahmedabad -- 380 009, India\\
              email: ashok@prl.ernet.in, orion@prl.ernet.in}
\date{Received  / Accepted }
\titlerunning{IR spectroscopy of V445 Puppis}
\maketitle

\begin{abstract}
$JHK$ spectroscopic and photometric observations are reported for the 
enigmatic, nova-like, variable V445 Puppis which erupted at the end of
  2000.
The near-IR spectra are hydrogen-deficient and unusually rich in CI lines. The 
important CI lines are found to be positioned at 1.133, 1.166, 1.175,
1.189, 1.26 and 1.689 ${\rm{\mu}}$m. 
Model calculations for the CI lines are done to make the identification 
of the lines  secure. Photometric data, taken on January 2.90 UT, 2001
 shortly after
the outburst, show the formation of an optically thin dust shell around V445 
Puppis. The  temperature and upper limit for the mass of the dust 
shell are estimated to be
1800K and $M{_{\rm dust}} \sim 10{^{\rm -9}}$$M$$_\odot$ 
respectively . A subsequent episode of massive dust formation 
 - indicated by the observed, deep visual dimming - is also seen
in $JHK$ images of early November 2001. V445 Puppis has shown two unusual 
properties for a nova 
i) the hydrogen-deficiency and He/C enrichment of the object as seen from 
optical and IR data and ii) synchrotron radio emission which was detected
nearly a year after it's outburst by other workers. The strange nature of the
object is discussed and it is shown that it is difficult to place it in the
known categories of eruptive variables viz. novae (classical, recurrent or 
symbiotic); born-again AGB stars; the new class of eruptive variables like
 V838 Mon;  and RCB/HdC stars. Tentative evidence for an accretion disk and 
 binarity of the V445 Puppis system is presented. It is debated whether
 V445 Puppis is a rare Helium nova or an unique object.


   \end{abstract}

\section{Introduction}
V445 Puppis - a nova like object - was first reported to be in outburst on 
30 December 2000 by Kanatsu (Kato and Kanatsu 2000).  The exact
 time of the of the object's eruption is not 
certain - an aspect that is  discussed in Sect. 3.1. Spectra in the
 visible region, taken in the early stages after the detection, showed many
permitted lines of FeII, CaI, CaII, OI and NaI (Fujii 2001; Liller 2001a, b;
 Shemmer et al. 2001; Wagner et al. 2001a, b). A striking feature of
the optical spectra was the absence of Hydrogen lines in the spectra. Near
IR spectra reported here also confirm the absence of Paschen and Brackett
Hydrogen lines in the $JHK$ bands. The deficiency of Hydrogen in V445 Puppis 
shows that it is not an usual nova and shows the  strangeness of the object.
This point is also made by Wagner et al. (2001a) who state that the
 optical spectrum of
V445 Puppis is not typical of classical FeII type novae, recurrent novae or
symbiotic novae. Mid IR (3-14 ${\rm{\mu}}$m) spectroscopy by Lynch 
et al. (2001) reveal  only a smooth featureless continuum deviating 
significantly from a black body distribution and which can be explained by
 invoking the presence of dust. Another interesting development in 
 the evolution of 
 V445 Puppis was the detection at radio wavelengths almost a year after
the outburst (Rupen et al. 2001a).  Radio emission has been seen in a few
 novae (Seaquist 1989) but such emission has the characteristics
 of thermal bremsstrahlung radiation emitted by ionized gas. In
 V445 Puppis, non-thermal synchrotron radiation is seen which is rather
 rare (another example of this being the old nova remnant GK Persei).  
  Thus V445 Puppis has all the hallmarks of being a very interesting object.\\
   
In this work, we  mainly present   spectroscopic results (and also some
photometric data) from $JHK$  observations of V445 Puppis  made
at five, fairly evenly-spaced epochs. These  should help in following the
temporal  evolution of V445 Puppis and understanding it's puzzling nature. 
From our results it would appear that V445 Puppis is a fairly unique object
and could be a Helium nova.  A Helium nova is believed to occur due to a
thermo-nuclear runaway on the surface of a degenerate white dwarf  accreting
Helium from it's helium-rich companion (Kato et al. 1989). The nova ejecta,
in such an outburst, is expected to be  hydrogen-deficient as found 
in V445 Puppis.

\begin{figure*}[t]
\centering
\includegraphics[bb=19 303 575 593,width=6.5in,height=3.2in,clip]{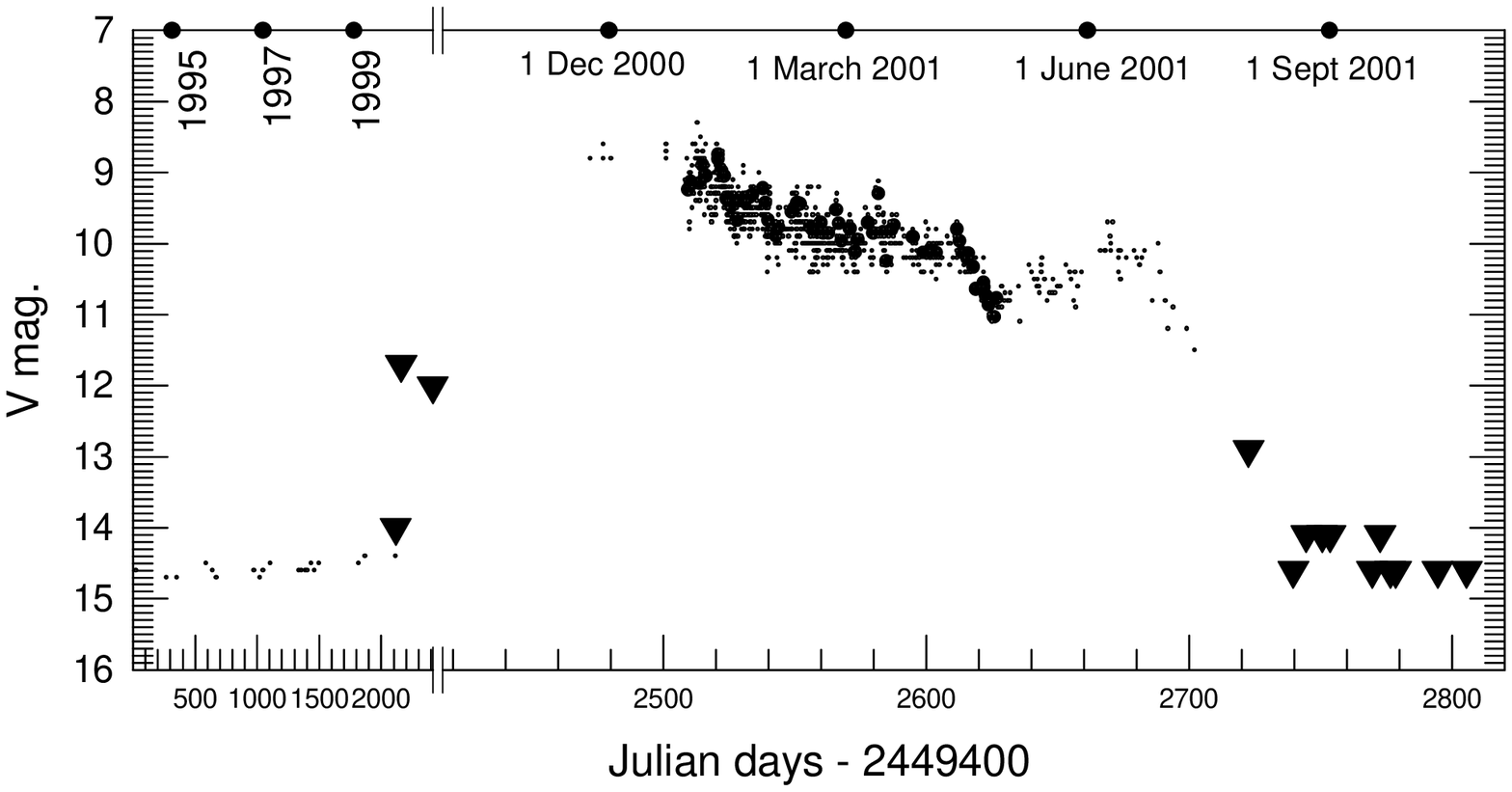}
\caption[]{ The $V$ band lightcurve of V445 Puppis between 1994 to
 2001 based on IAU circulars and VSNET reports.The small dots 
represent visual/photographic estimates and the circles represent 
photo-electric/CCD values. The triangles give upper limits on the magnitudes.}
\label{fig1}
\end{figure*}
  
\section{Observations}

 Near-IR $JHK$ spectra at a resolution of $\sim$ 1000 were obtained at the
 Mt. Abu 1.2m  telescope using a Near Infrared Imager/Spectrometer
 with a 256$\times$256 HgCdTe NICMOS3 array. We present here the spectroscopic
 observations  of five days viz.  January 1.63 UT, January 20.81 UT,
  February 1.73 UT,  February 18.70 UT
 and  March 1.66 UT, 2001.   
 In each of the $J$, $H$ and $K$ bands a set of at least two spectra were
  taken. In each set the  star was offset to two 
 different positions of the  slit (slit width $=$ 2 arc sec.\ ). The 
 signal to noise ratio of the spectra, as determined using IRAF, is moderate
  and ranges between 30 to 60 in the $JHK$ bands.
 The exposure times for the spectra were typically between 60 to 120 seconds.
 Spectral calibration   was done using the OH sky lines 
 that register with the spectra.  The comparison star  HR 2988 was used for
 ratioing the spectra in all cases.  We believe the telluric
  lines, present in the spectra, have been well removed in the process of
  ratioing. This is because of i) the  close proximity of
 HR 2988 to V445 Puppis and ii) also the small time gap (typically 10-15 
 minutes)
between observing the spectra of V445 Puppis and HR 2988 in any of 
 the
 $JHK$ bands. This ensures that
both the stars were at a similar airmass.
 The ratioing process should therefore
remove the telluric lines reasonably 
 well.\\

 Photometry in the $JHK$ bands was  performed on  January 2.90 UT and
   November 1.96 UT, 2001 using
  the NICMOS3  array, mentioned above, in the imaging mode.
   The sky was photometric on both days with a typical seeing of
   1 arc second. Several frames  in 4 dithered positions, offset
   from each other by  30 arcsec, were obtained in all the filters. The
  times for all the individual frames was 100 ms on 2 Jan. 01 and
   ranged between 30 to 60s on 1 Nov. 01.  The total integration time
   in each of the dithered positions are as follows i) 400 ms in each of the
    $JHK$ bands for 2 Jan.01 and ii) 3 minutes in $J$ and 90s in $H$
    on 1 Nov. 01.
 The sky frames were generated using these dithered frames. The mean
air-mass at the time of observations for V445 Puppis was
 1.79 on 2 Jan. 01 and 
1.59 on 1 Nov. 01. The UKIRT standard star HD 77281 was used for photometric
calibration  on 2 Jan. 01 and observed soon after V838 Mon. As a 
crosscheck, HR 2956 ( $V$= 6.50, spectral type B7V) was also 
observed. The adopted $JHK$
magnitudes for HR 2956 were taken from the 2MASS survey to be 
$J$ = 6.642, $H$ = 6.723 and $K$ = 6.686. For the 1 Nov 01 observations, FS 13 
($J$ = 10.517, $H$ = 10.189 and $K$ = 10.137) was used as the standard star
 (Hunt et al. 1998). The atmospheric extinction corrections were done assuming 
average values of $k$$_{\rm J}$ = 0.15, $k$$_{\rm H}$ = 0.15 and
$k$$_{\rm K}$ = 0.1 mag for the Mt. Abu Observatory site.  The
results from the photometry are presented later in Table 2.  The 
near-IR  photometric
 and spectroscopic data were all reduced using IRAF.

\begin{figure}[t]
\centering
\includegraphics[bb=58 180 368 603,width=3.2in,height=4.5in,clip]{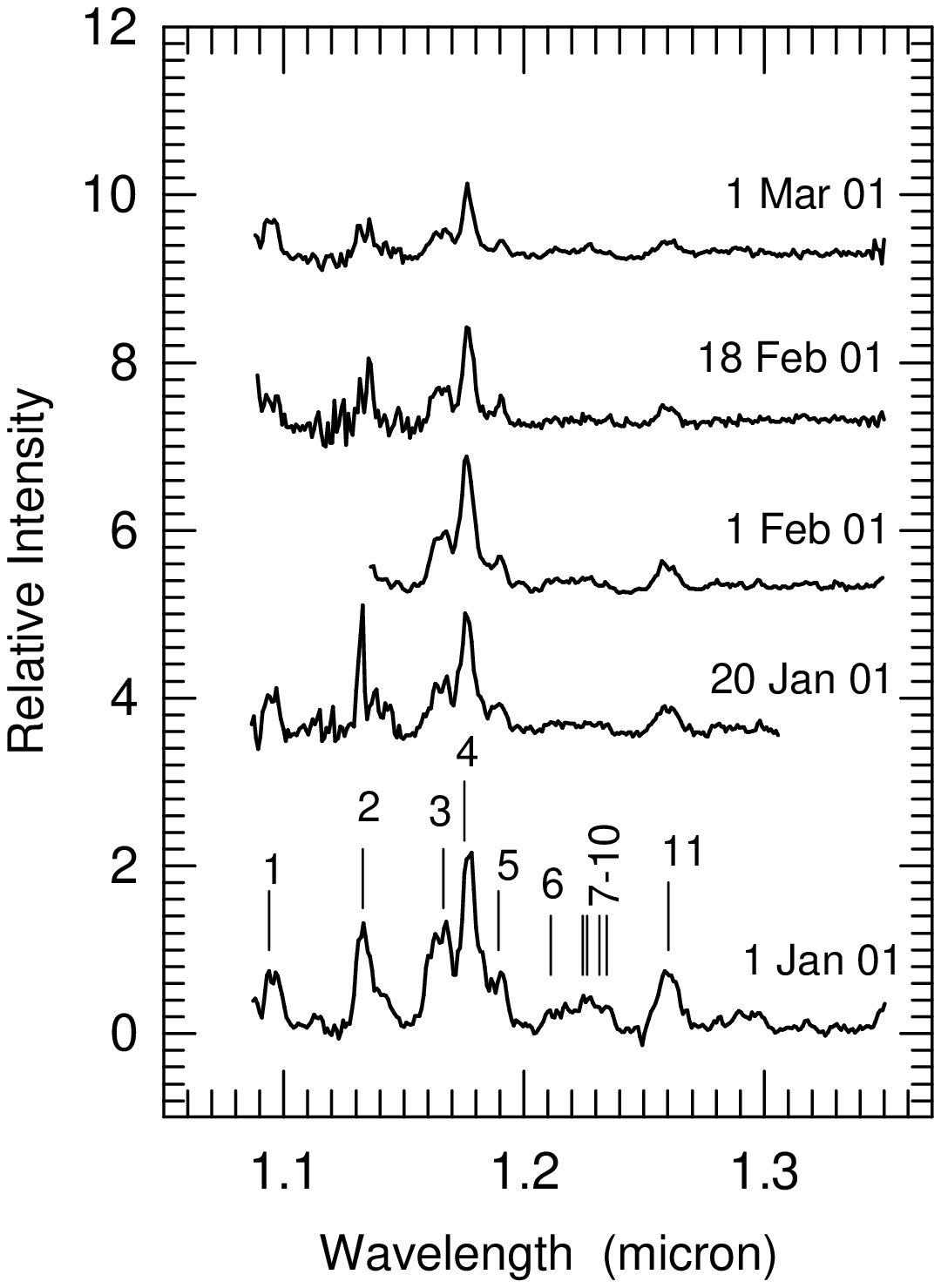}
\caption[]{ The $J$ band spectra of V445 Puppis are shown at different epochs.
The spectra have been offset from each other for clarity. The identified lines
 are
 marked and numbered - their details are given in Table 1.}
\label{fig2}
\end{figure}


\begin{figure}
\centering
\includegraphics[bb=69 181 368 603,width=3.2in,height=4.5in,clip]{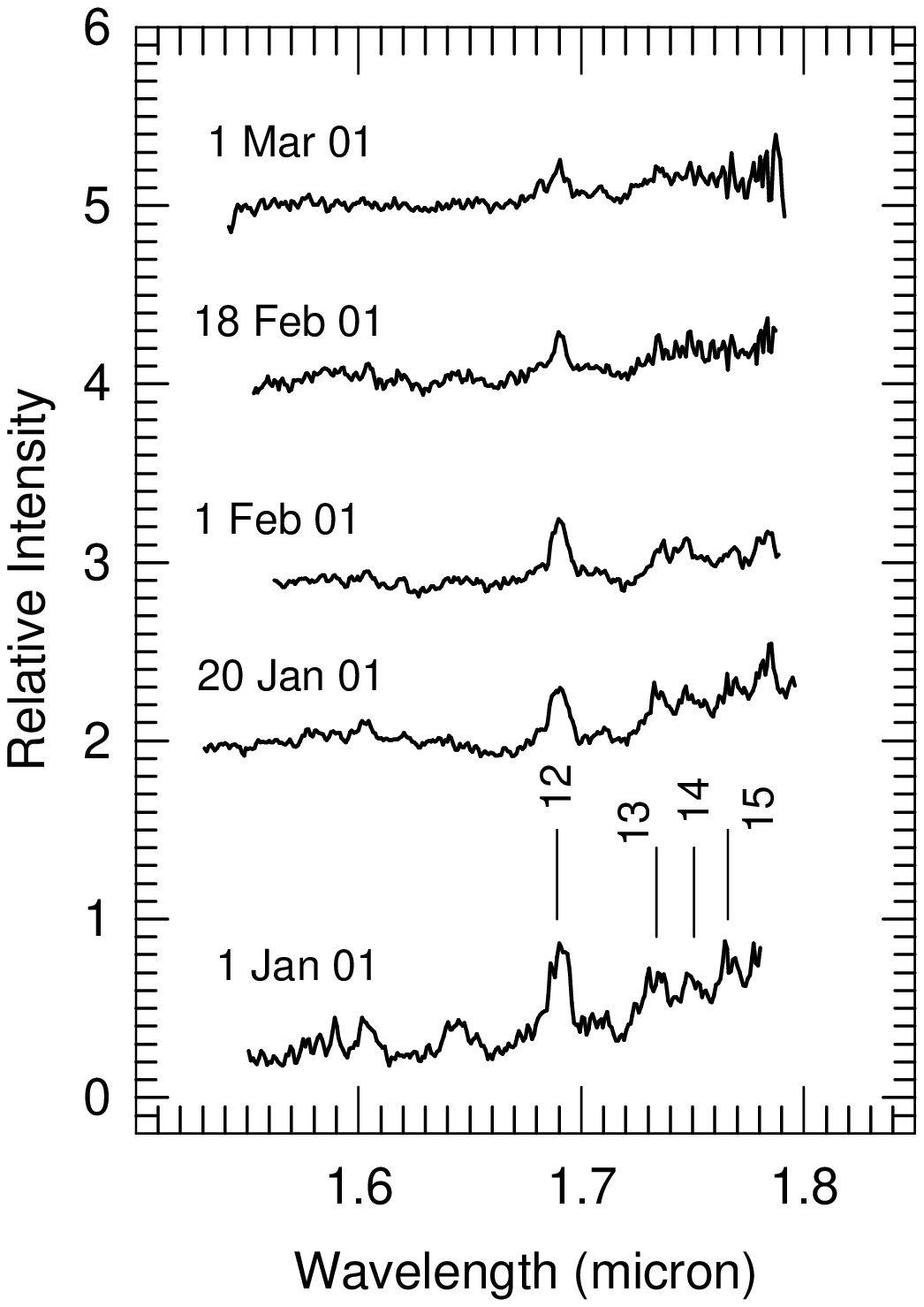}
\caption[]{The $H$ band spectra of V838 Mon are shown at different epochs. 
The spectra have been offset from each other for clarity. 
The identified lines are numbered and marked - their details are given
 in Table 1}
\label{fig3}
\end{figure}

\section{Results}
\subsection{The Light Curve of V445 Puppis}
 The outburst date of V445 Puppis is uncertain but can be constrained from
 the VSNET reports. This part of the sky has been monitored by Takamizawa 
 between March 1994 to  December 1999 and he does not find any object brighter
 than 14 magnitude on his photographic plates (IAUC 7552). 
 Subsequent photographic records from VSNET
(\protect  http://www. kusastro.kyoto-u.ac.jp.vsnet) show that the object
 was fainter than  12th magnitude
on 26 September 2000 and had brightened to 8.8 magnitude by 23 November 2000. 
Thus the outburst occured between 26 Sept. to 23 Nov. 2000. Lynch et al.
(2001) have 
inadvertently concluded that the outburst occured between 3 to 31 December 
2000.   The progenitor of V445 Puppis has been identified by Platais et al.
(2001) as a 13.6 magnitude star (identification number 22727221) in the
U.S. Naval Observatory CCD Astrograph Catalog (UCAC1). In the  
the USNO A2.0 catalog, the corresponding identification
number is 0600-06937901 with the blue and red magnitudes 
listed as 13.8 and 13.2 respectively.
Fig. 1 shows the evolution of the light curve for V445 Puppis. While the
 decline of the light curve is similar to that of a slow nova, the outburst
amplitude of approximately six magnitudes is small for a classical nova. Other
differences between the object and  classical novae are discussed in Sect. 4.

\begin{figure}[t]
\centering
\includegraphics[bb=68 179 368 603,width=3.2in,height=4.5in,clip]{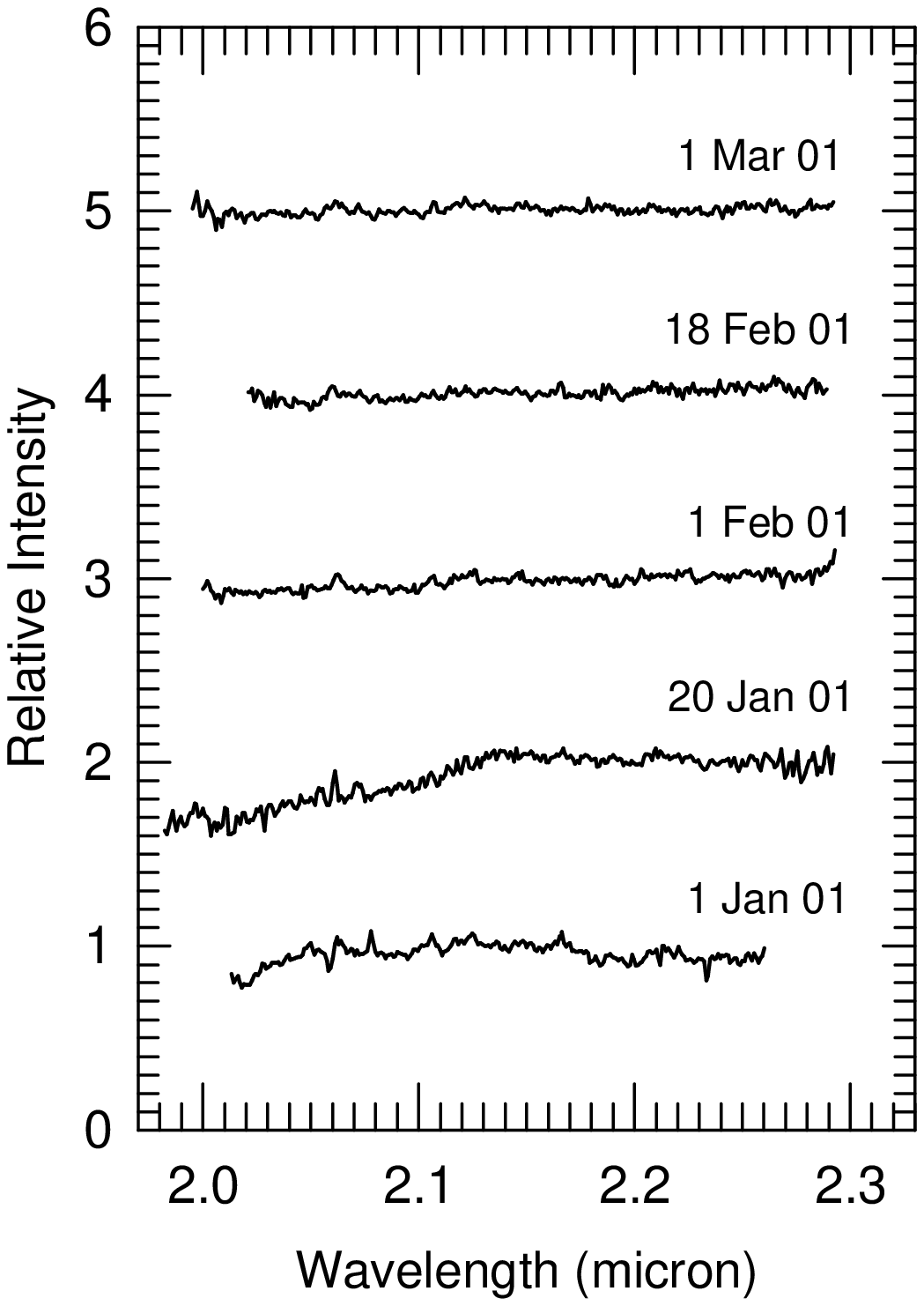}
\caption[]{The $K$ band spectra of V838 Mon are shown at different epochs. 
The spectra have been offset from each other for clarity.}
\label{fig4}
\end{figure}

 \subsection{$JHK$ spectroscopy}
The main thrust of this work has been in the $JHK$ spectroscopy of V445 Puppis.
The object was observed at regular intervals between 1 Jan. to 1 March 2001. 
The $JHK$ spectra are 
shown in Figs. 2, 3 and 4. Of these, the more interesting spectra are in 
the $J$ band where several emission lines can be seen. The emission lines that
 are seen in the $J$ and $H$ spectra are numbered in Figs. 2 and 3. 
 Their identification and other relevant details, are listed in Table 1. 
Almost all the lines that are present appear to be due to neutral Carbon.
The identification of the CI lines is done on two bases. First we have 
compared the strengths and positions of these lines with the laboratory IR
spectra of CI as given by Johansson \& Litzen (1965) and Johansson (1966)
 and a good match is found. Furthermore,  we have
computed a simple model spectrum that should arise from CI emission from
the shell. For this spectrum, we have selected 170 of the 
strongest CI lines between 1.08 and 2.5 ${\rm{\mu}}$m 
with their transition probabilities, from the atomic line list compiled by 
Kurucz (\protect http://kurucz.harvard.edu/linelists.html).  The strength 
of any line is computed as follows.
If the line originates in a downward transition from level 2 to 1 then,
the luminosity of the line  $L$$_{\rm line}$ is  given by:
 
 \begin{equation}
 L{_{\rm line}} = N{_{\rm 2}}A{_{\rm 21}}h \nu V  
  \end{equation}
    \vskip 2mm
  where $N$$_{\rm 2}$ is the number density  of CI atoms in the
 excited state  2,   
  $A$$_{\rm 21}$ is the Einstein  coefficient of spontaneous emission,
  $h$$\nu$ is the energy of the emitted photon  and $V$ is the volume of the
  emitting gas.\\

From Eq. (1), it is seen that, the strength of any line will essentially
 depend only on $A$$_{\rm 21}$  and $N$$_{\rm 2}$. If the populations of the
  upper level ($N$$_{\rm 2}$) for the
different CI lines are temporarily assumed to be the same, then the line 
strength depends basically  only on  $A$$_{\rm 21}$. Knowing the transition
probabilities, we have computed the relative line strengths from Eq. (1).
We assume that the shape of each line can be reasonably represented by a
Gaussian whose FWHM has been chosen to be approximately 1500 km/s - 
representative of the
observed linewidths in V445 Puppis. Co-addition of all the Gaussians - 
corresponding to all the lines- yields a model spectrum which is shown 
in Fig. 5 (labeled as model 1). The observed $J$ band spectrum of 1 Jan. 01,
 in which the emission lines are most prominent, is also shown in the same
  figure  for comparison.  A more accurate
calculation has taken into account the population of the upper levels
 $N$$_{\rm 2}$ in calculating the relative line strengths. Assuming the
  emitting gas to be in local thermal equilibrium, $N$$_{\rm 2}$ will be given 
  by a Boltzmann distribution viz.

   \begin{equation}
   N{_{\rm 2}}/N(C) = (g{_{\rm 2}}/U)e{^{(\rm -\chi{_{02}}/kT)}}
   \end{equation}
   \vskip 2mm
   where $U$ is the partition function, $T$ is the    temperature of the gas, 
   $\chi$${_{\rm 02}}$ is the energy difference  between the ground state
    and level 2, $g$$_{\rm 2}$ is the statistical weight of the upper level
     and $N(C)$ is
 the total number density of the Carbon atoms. The energy level values for the 
 different states (to compute $\chi$$_{\rm 02}$) and $g$$_{\rm 2}$ values  have
 been taken from  Kurucz's line list. The partition function values, for
  different temperatures, have been taken from Allen (1976) and Aller (1963).\\

\begin{figure}[t]
\centering
\includegraphics[bb=170 200 429 558,width=3.2in,clip]{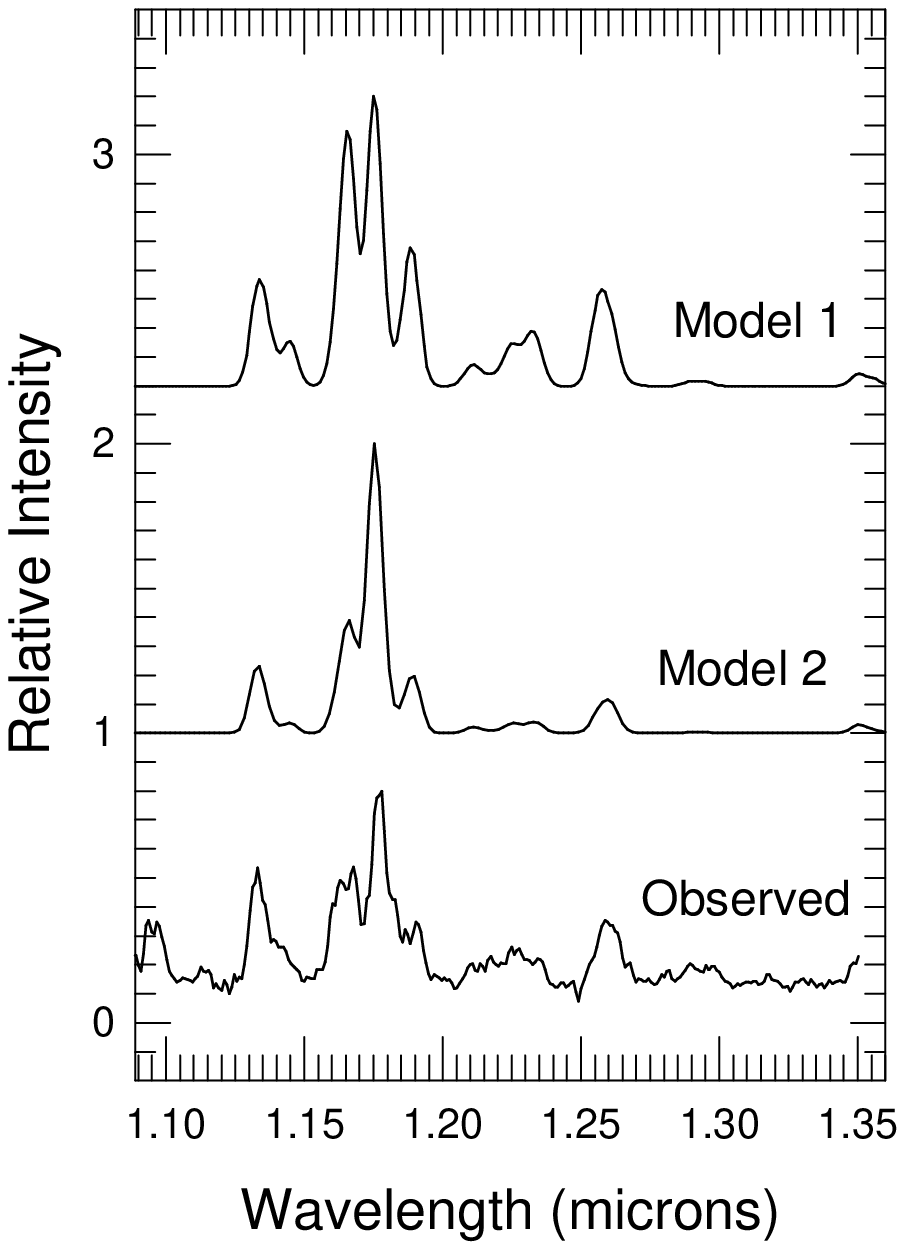}
\caption[]{Model spectra compared with the $J$ band spectra of 
1 Jan. 01. The
details are available in the text of Sect. 3.2.}
\label{fig5}
\end{figure}

Nova shell ejecta  have  temperatures typically in the  range 4000-5000K
(Williams 1994). However we find slightly higher temperatures reproduce the
observed spectra better.    
In particular, the model spectrum computed for T = 8000K (labeled as model 2)
 is
also shown in Fig. 5. Changes in the adopted temperature do change the
 relative 
strength
of the different lines,but marginally, and not in any
  drastic fashion. 
 However, the primary purpose of the simulated spectra
of Fig. 5  is only to secure the identification of the CI
lines. Fig. 5 only shows graphically the expected positions of the CI lines and
gives a rough simulation of their expected strengths. As may be seen all the 
strong CI lines  are reproduced, and to a large extent, even the weaker lines.
Based on the results of Fig. 5, it would appear 
reasonable to identify most of the observed lines in the $J$ band with CI.
The model spectra for $H$ and $K$ have also been computed but not shown here.
The model results for $H$ band show the strongest  CI line to be at 
1.69 ${\rm{\mu}}$m which matches what is observed at that wavelength. A few
 other lines between 1.7 to 1.8 ${\rm{\mu}}$m are predicted as well ( weaker
  than the
  1.69 ${\rm{\mu}}$m line). Their positions have been marked in the $H$ band
 spectra of Fig. 3 and these lines are also  seen in our data. In the $K$ band 
no line of any significant strength is predicted. This agrees with the rather
 featureless $K$ band spectra that we observe (Fig. 4). \\

                In Table 1, we have given the equivalent
        widths for the emission features seen in the spectra
        of 1 Jan. 01 where the lines are the
        strongest. Since some of the features are blended, we have given
        the combined equivalent width of the blend. In particular,
        combined equivalent width's have been given in Table 1 for
        features 3 to 5 and 6 to 10 which are blended. The reality of the
        features can be judged by comparing their equivalent widths with
        those of modulations in the continuum due to noise. The
        equivalent width values for 
        such modulations are in the range of 3 to 7$\AA$ in the $J$ and
        $H$ band spectra. In comparison, on 1 March 2001, when the lines
        were the weakest, the equivalent widths for the features which are
        clearly discernible viz. 1, 2 , 3-5, 6-10, 11 and 12 are
        25, 39, 102, 27, 22 and 41$\AA$  respectively. The noise in the
        continuum has similar equivalent widths as before. In general, the
        $H$ band features are a little noisy except for the 1.689 ${\rm{\mu}}$m
        line.
\\
               
                While it appears that CI is the main contributor to
        the $JHK$ spectra, it is possible that there is some contribution
        from CII also. This is specially so because CII lines have been seen in
        the optical spectra as will be discussed shortly.  We have
        determined some of the stronger CII lines ( based primarily on
        their transition probabilities) in the $JHK$ region from Kurucz's
        line list. It will need a detailed model - beyond the scope of
        this work - to calculate the absolute strengths of these CII lines
        vis-a-vis the CI lines. But their expected wavelength positions
        should give some indication whether they are present in the $JHK$
        spectra or not.  The expected CII lines are found to lie at
        wavelengths of 1.093, 1.144, 1.158, 1.674, 1.700 and 2.222
        ${\rm{\mu}}$m. There is a line at 1.09 ${\rm{\mu}}$m which is
         consistently seen in the
observed spectra - this could be due to
          CII. Out of the other CII
        lines the 1.144, 1.158 and 1.700 ${\rm{\mu}}$m lines may be present
         in the
observed spectra. The 1.144 and 1.158 ${\rm{\mu}}$m CII 
         lines could be lying on
the wings of the observed 1.133 and 
         1.165 ${\rm{\mu}}$m CI lines - giving them a
 slightly broader 
         appearance.  Similarly, the expected 1.700 ${\rm{\mu}}$m CII
        line may be blended with the 1.689 ${\rm{\mu}}$m CI line.  Thus
         there is some
evidence for the presence of CII in the $JHK$ spectra.

               It must be mentioned that in an earlier IAU circular (Ashok
        \& Banerjee 2001) we had attributed some of the
        CI lines reported here to CIII and CIV. This
        was because some CIII and CIV lines do occur at similar
        wavelengths as CI - for e.g. as seen in the spectra of Wolf-Rayet
        stars (Eenens et al.  1991 ). But some of these strong CIII and CIV
        lines are not seen here.  Furthermore, since lines from low-ionization
        species have been definitely identified in the optical spectra of
        V445 Puppis, it is more consistent to attribute the observed
        emission lines to CI ( with some contribution from CII possibly),
        rather than high-ionization species like CIII or CIV.

               It may be seen from the $JHK$ spectra that V445 Pup is
        hydrogen-deficient. In the $K$ band there is no sign of the
        Brackett gamma (Br$\gamma$) line at 2.1656 ${\rm{\mu}}$m.
        Similarly other Brackett series lines (Brackett 10 to 19) which
        are commonly seen in the $H$ band in the spectra of classical
        novae are missing. Paschen beta (Pa$\beta$) at 1.2818
        ${\rm{\mu}}$m is also not present in the $J$ band spectrum.  The
        only possible indication for the presence of Hydrogen is the 1.09
        ${\rm{\mu}}$m line which coincides with Paschen gamma
        (Pa$\gamma$). However, Pa$\gamma$ is expected to be weaker than
        Pa$\beta$ which, in the present data, is completely absent.
        Hence it is difficult to attribute this line to Hydrogen
        Pa$\gamma$  and it is  more likely to be due to CII as
        discussed earlier.\\
        
                We do not find the signature of Helium lines in the $JHK$
        spectra. However, the presence of He is more conclusively seen in
        the optical spectra wherein several HeI lines are prominently seen
        (Wagner et al. 2001c) while at the same time the
        hydrogen-deficiency is manifested by the weakness/absence of the
        Balmer Hydrogen lines.  It is also important to note that many
        Carbon lines are also strongly seen in the optical spectra (Kamath
	\& Anupama 2002; Wagner's site at
        http://vela.as.arizona.edu/$\sim$rmw /v445pup.html) at different
        epochs after the outburst. In fact in a sample spectrum by M.
        Fujii, Bisei Observatory, taken soon after the outburst, the
        strongest optical line that is seen is the CII 6582$\AA$ line
        (http://www1.harenet.ne.jp/~aikow/0113pnp.gif). The Fujii spectrum
        also shows the presence of the HeI lines. Thus it would be fair to
        qualitatively say that the IR and optical evidence indicate that
        not only is V445 Puppis a hydrogen-deficient object but also that
        it is rich in Carbon and Helium.

 \begin{table}
 \caption{A list of the lines identified from the $J$ \& $H$ spectra shown
 in Figs. 2 and 3.  Please refer to the text for greater details on the
  equivalent widths listed below for 1 January 2001. (u.i = unidentified)}
\begin{tabular}{lllllll}
\hline \\ 
Line \#  & $\lambda$(${\rm{\mu}}$m) & Species & Other & Eq.\\
(From Figs &  &  & Contri- & Width\\
2 \& 3) &  &  & butors & ($\AA$)\\
\hline 
\hline \\ 
1  & 1.0938   & CII &       & 34  \\
2  & 1.1330   & CI  &   CII & 110 \\
3  & 1.1658,  & CI  &   CII & 320 \\
   & 1.1669   & "   &       &     \\
4  & 1.1748,  & CI  &       &     \\
   & 1.1753,  & "   &       &     \\
   & 1.1755   & "   &       &     \\
5  & 1.1895   & CI  &       &     \\
6  & 1.2112   & CI  &       & 69  \\
7  & 1.2244,  & CI  &       &     \\
   & 1.2248   & "   &       &     \\
8  & 1.2264   & CI  &       &     \\
9  & 1.2314   & CI  &       &     \\
10 & 1.2336,  & CI  &       &     \\
   & 1.2348   & "   &       &     \\
11 & 1.2581,  & CI  &       & 65  \\
   & 1.2601,  & "   &       &     \\
   & 1.2614   & "   &       &     \\
12 & 1.6890   & CI  &   CII & 53  \\
13 & 1.7338   & CI  &       & 15  \\
14 & 1.7505   & CI  &       & 9   \\
15 & 1.7659   & u.i &       & 12  \\
  
\hline
\end{tabular} 
\end{table}

 \subsection{Distance and Reddening}
We briefly discuss the distance and reddening towards V445 Pup since these 
parameters enter the photometric calculations described in the coming
subsection. The distance to V445 Pup is poorly known. The only estimate for 
the distance $d$ is by Wagner who give an upper limit of 3 kpc based on the
strength of an interstellar absorption line at 5780$\AA$
(http://vela.as.arizona.edu/$\sim$rmw/v445pup.html). For such an adopted upper 
limit for $d$, the 
extinction maps by Neckel et al. (1980) give a value of $E(B-V)$ $\sim$ 0.25
 in the 
direction of V445 Puppis. This value is derived by converting Neckel et al.'s 
(1980) values of $A{_{\rm v}}$ using the relation 
  $A$${_{\rm V}}$ $=$ 3.1$E(B-V)$ (Koornneef 1983).
As an additional confirmation for $E(B-V)$, we also
used  UBV photometric data for 110 stars by Wooden (1970) in an
approximately  5$\times$5 degree field around V445 Puppis. Since the spectral
 type is known, and hence the intrinsic $(B-V)$$_{\rm 0}$ color, the 
 excess color  $E(B-V)$ can be found from the observed $(B-V)$ color.  
 $A$$_{\rm v}$ can hence be found. The distance can then be determined by the 
 standard relation 
 $m{_{\rm v}}$ - $M{_{\rm V}}$ = 5log$d$ - 5 + $A{_{\rm v}}$ . The values
  for the
  intrinsic $(B-V)$$_{\rm 0}$ colors and absolute magnitudes, for
  different spectral types, were taken
from Lang (1991) and Allen (1976). We have plotted in Fig. 6 the  $A{_{\rm v}}$
versus 
distance relation based on the Wooden (1970) data.  Since
 a 5$\times$5 degree 
field is rather large and extinction can vary considerably along any line of
sight in it, we have also shown in Fig. 6 the extinction for stars within a 
smaller, 1 degree field  of V445 Puppis. Based on the results of
 Fig. 6 and also on the Neckel et al. (1980) data, we feel it reasonable to
 adopt a value of $E(B-V)$ = 0.25 for V445 Puppis, unless the distance to it
 is much greater than 3 kpc. It may be noted, that the extinction in this 
 direction is generally low.

\begin{figure}[h]
\centering
\includegraphics[bb=58 367 337 602,width=3.1in,clip]{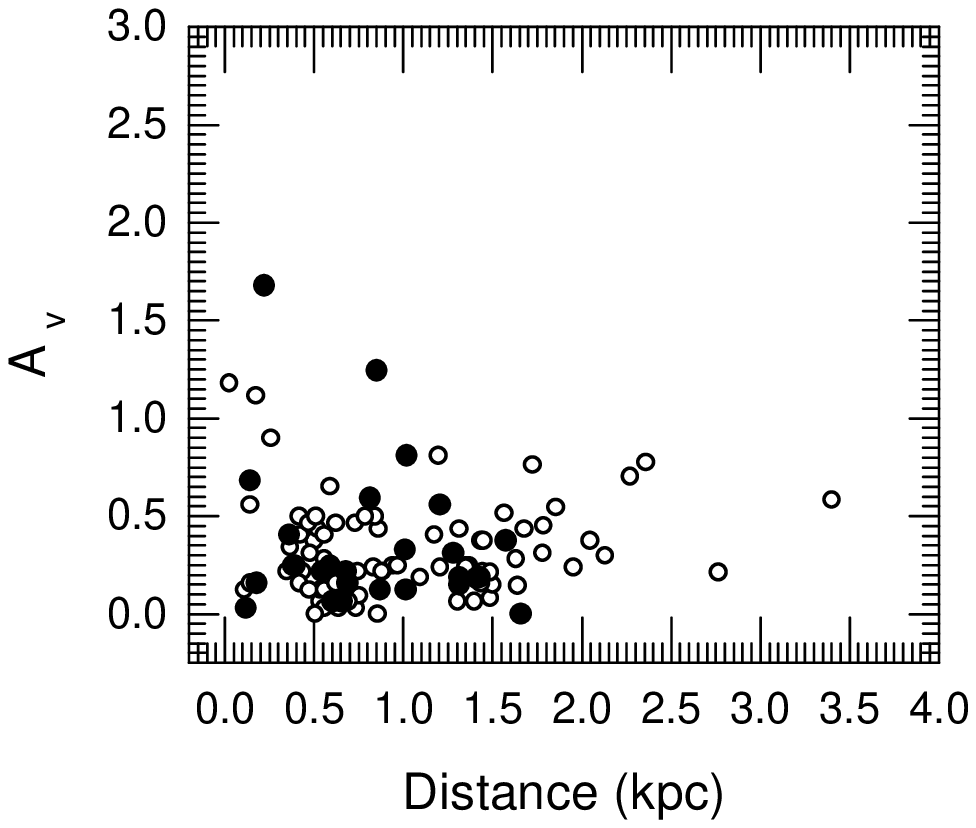}
\caption[]{ An $A$$_{\rm v}$ versus distance plot for stars in a 5$\times$5 degree
 field around V445 Puppis based on data from Wooden (1970). The stars within
 a smaller, 1 degree field of V445 Puppis are plotted by
 filled circles, the other stars by empty circles. }
\label{fig6}
\end{figure}

\subsection{Near Infrared photometry }
The details and results of the photometry are given in Table 2.
 The $JHK$ magnitudes listed are the observed magnitudes. 
 We have used Koornneef's (1983) relations viz. 
  $A$${_{\rm V}}$ $=$ 3.1$E(B-V)$, 
 $A$${_{\rm J}}$ $=$ 0.265 $A$${_{\rm V}}$,
  $A$${_{\rm H}}$ $=$ 0.155 $A$${_{\rm V}}$ and
 $A$${_{\rm K}}$ $=$ 0.090 $A$${_{\rm V}}$ to correct for interstellar
 extinction.
  Absolute flux calibration was done by adopting zero magnitude
  fluxes from Koornneef (1983).
\begin{table}[h]
\caption[]{$JHK$ photometry of V445 Puppis}
\begin{tabular}{llll}
\hline\\
Obs. date (UT)&  $J$ & $H$  & $K$ \\
\hline 
\hline \\ 
 2.90 Jan 2001& 6.35 $\pm$ 0.06   & 5.20 $\pm$ 0.03  & 4.41 $\pm$  0.12  \\
 1.96 Nov 2001& $>$ 15.5    & 12.6 $\pm$ 0.20  & - \\

\hline
\end{tabular} 
\end{table}
We have plotted the spectral energy distribution (SED) of V445 Puppis for 
2 Jan. 01, after
correcting for $E(B-V)$ = 0.25, 
in the upper panel of Fig. 7. In this figure 
the $JHK$ fluxes are from the 
present work 
whereas the UBVR values are taken from Gilmore (2001). It may be
noted that Gilmore's (2001) observations are for Jan. 6.58 UT, 2001 and slightly 
separated in 
time from our 2 Jan. 01 $JHK$ observations. However, since both sets of 
observations are almost quasi-simultaneous, no significant  error is expected to
be introduced in the results of Fig. 7. This is especially so, since as seen
from the light curve of Fig. 1, V445 Puppis was not showing  strong photometric
changes around this time.\\

 The SED  of 2 Jan. 01  could not be fit by a single
black body curve as it shows a significant IR excess. Accordingly it has been 
fit 
by a 7500K black body component ( corresponding to V455 Puppis proper) and a 
cooler component of temperature 1800K. The cooler component arises from dust
that has formed around V445 Puppis.   The presence of dust could give an
additional extinction apart from the interstellar extinction corresponding to 
the adopted value of $E(B-V) = 0.25$. In view of this we have calculated 
various combinations of the SED that result for variations of $E(B-V)$ from
        0.25 to 0.5; the hot blackbody component  from 6000 to 12000K
        and the cool blackbody component (dust) from 1200 - 2200K. We find
        that $E(B-V)$ $\sim$ 0.25 still continues to give an optimal fit to
         the data
        indicating that there is no significant extinction from the dust. 
              We also find that almost equally good fits to the data
        are found for a hot component temperature in the range
        7500-8500K combined with dust temperature in the range 1750-1850K.
        We believe a dust temperature of 1800K is fairly representative 
        of the observed data and
        adopt this value for subsequent calculations for the mass of the
         dust shell.\\

 Lynch
et al. (2001) have also found evidence for the 
presence of dust
  from their 3-14 ${\rm{\mu}}$m  $LMN$ band spectroscopy of 31 January 2001.
 However they conclude that the dust, most likely, pre-existed before the 
 outburst rather
than  created post-outburst. The premise for this inference is that
dust could not have formed so early after the outburst. But, as mentioned 
earlier in section, Lynch et al. (2001) have placed the outburst date a month
 later than  what we believe to be correct. In classical novae, the  time scale
  for dust to form is typically 50 - 70 days after outburst (Gehrz 1988). 
  However there are cases like V838 Her 
 (Chandrasekhar et al. 1992 and references therein) when dust  formed as 
 early as eight 
 days after outburst. In the case of V445 Puppis, we feel that sufficient time
 after the outburst may have elapsed for dust to form  and the dust that is 
 seen here has been created post-outburst. This view is bolstered by looking
  at the  pre-outburst SED of V445 Puppis shown in the lower panel of Fig. 7.
   Here, the pre-outburst fluxes are  given by the filled circles which are
derived from the  $B$ \& $R$ magnitudes ( from the   USNO A2.0 database) and the
 $JHK$ magnitudes from the 2MASS survey. As may be noted
 there is no sign of a pronounced IR excess in the   pre-outburst SED of 
 V445 Puppis. Discussion of the other plots in the lower panel of Fig. 7 is 
 deferred to Sect. 4.\\
 
 The mass of the dust shell can be inferred from the infrared excess that is 
 seen in the data of 2 Jan. 01. Following Woodward et al. (1993), 
  the  mass of the dust shell (in units of $M$$_\odot$) is given by
  
 \begin{equation}
 M{_{\rm dust}} =1.1\times 10{^{\rm 6}} (\lambda F{_{\rm \lambda}}){_{\rm max}}d{^{\rm 2}}/ T{_{\rm dust}}{^{\rm 6}}
  \end{equation}
    \vskip 2mm
 
 In the above equation, $T{_{\rm dust}}$ is the black body temperature of the 
 dust in units of $10{^{\rm 3}}$K, $d$ is the distance to the object in kpc and 
 $(\lambda F{_{\rm \lambda}}){_{\rm max}}$
 is the flux, in Wcm${^{\rm -2}}$, measured at the peak of the SED for the 
 dust. Eq. (3) assumes the dust is composed of Carbon particles of size $\le$
 1${\rm{\mu}}$m and having a density 2.25 gm/cm${^{\rm 3}}$.
  The assumption that the dust is made of carbon particles may be quite
 valid in the case of V445 Puppis because of the strong presence of Carbon
  in the optical/IR spectra. Further, the mid - IR spectra of
        Lynch et al. (2001) show a featureless continuum which is generally
        attributable to dust in the form of carbon/graphite.
 From the data for 2 Jan. 01 we get
$M{_{\rm dust}} = 1.8 \times 10{^{\rm -10}}d{^{\rm 2}}$ for
 $T{_{\rm dust}}$ = 1800K and $(\lambda F{_{\rm \lambda}}){_{\rm max}}
= 1.716 \times 10{^{\rm -15}}$ Wcm${^{\rm -2}}$.  For a maximum 
 value of $d$ = 3 kpc, the upper limit of the mass of the dust shell is
 found to be   $M{_{\rm dust}} = 1.62 \times 10{^{\rm -9}}$$M$$_\odot$.\\

\begin{figure}[h]
\centering
\includegraphics[bb=67 125 497 535, width=3.5in,clip]{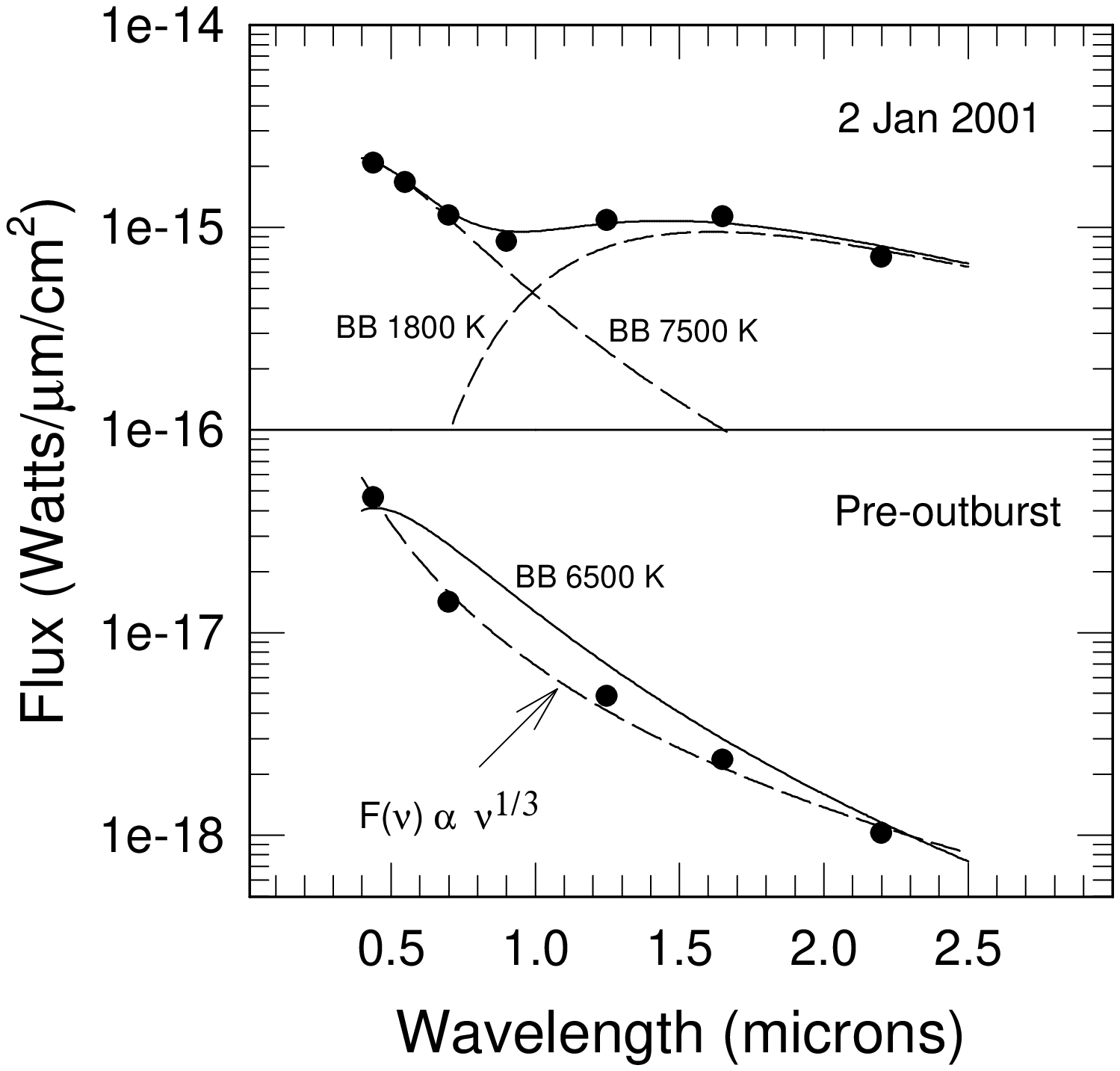}
\caption[]{ The post-outburst energy distribution of V445 Puppis
 on 2 Jan. 01
is shown in the top panel. It has been matched by the 
 sum of 2 black body 
curves - one at 7500K and the other for a dust component at 1800K. The bottom
panel shows the pre-outburst flux distribution based on 2MASS data. A 6500K 
black body fit is shown to this and also the expected distribution from an 
accretion disk following a $F{_{\rm \nu}}$ $\alpha$ $\nu{^{\rm 1/3}}$ relation.
 For greater details see text.}
\label{fig7}
\end{figure}
 
Regarding the photometric data for 1 Nov 01, an accurate estimate could not 
be made of the $K$ band magnitude because of inadequate S/N in the images.
However the object is clearly seen in the $K$ band, and also in the $H$ band, 
whereas it is not visible in the $J$ band images. This 
 can be seen from Fig. 8  showing the $J$ and $H$ band images (which have
  lost a little in reproduction). The position of V445 Pup is circled.
  Due to its absence in the $J$ band,  only a lower limit could be put
 for the $J$ magnitude by
 comparing the faintest, detected star in our $J$ image with its corresponding
 2MASS $J$ magnitude.  The limiting magnitude of the $J$ band
  image of  1 Nov. 2001 is 15.5. Our $JHK$ images/magnitudes confirm that
 the object was shrouded in  thick dust shell by then. 
 A similar conclusion was drawn earlier by Henden et al. (2001) based on 
observations of September- October 01. Henden et al. (2001) find that the
 star is not detectable  in the 
 $V$ and $I$ bands down to  limiting magnitudes of 20 and 19.5 
 respectively - in the $K'$ band 
 the reported magnitude is 9.15. The light curve of Fig. 1 also shows the 
 steep decline in brightness, at around this time, indicating the
 onset of a heavy dust formation phase. 

\begin{figure}[h]
\centering
\includegraphics[bb=36 184 308 608, width=3.5in,clip]{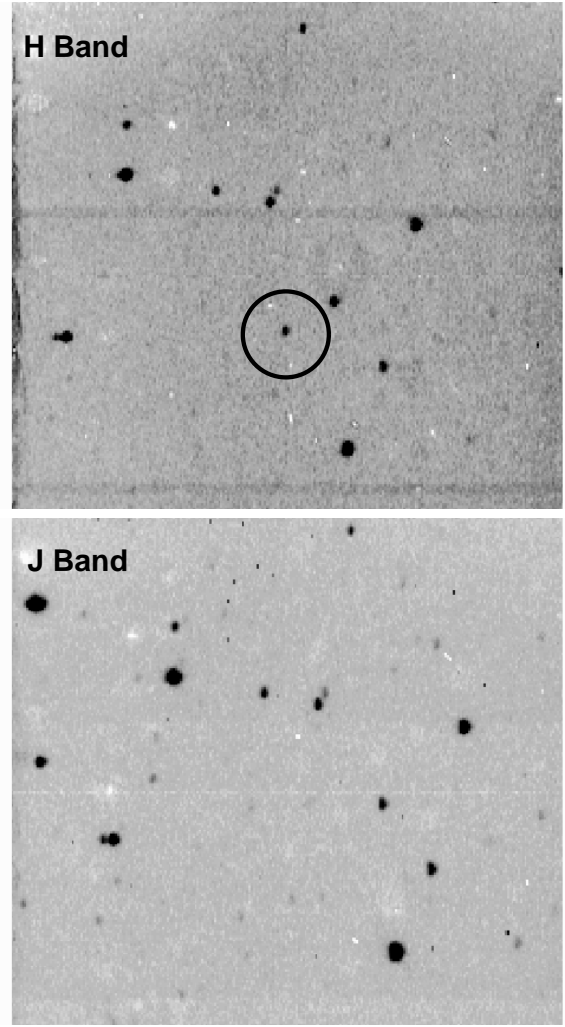}
\caption[]{ $J$ and $H$ band images (4$\times$4 arc. min) of V445 Puppis of 
1 Nov. 2001.  The object,
 which is circled in the $H$ band image, is not seen
 in the $J$ band. North is on the top and East is to the right.}
\label{fig8}
\end{figure}

\section{Discussion: The nature of V445 Puppis}       
 The nature of V445 Puppis is rather enigmatic. Among eruptive variables that
  show similar outbursts, one may first consider the novae. V445 Puppis
   differs from a classical 
 novae in the amplitude of it's outburst and also the deficiency of Hydrogen 
 in its 
 spectrum. The strong presence of a large number of Carbon lines in the IR 
 spectra is again unusual for a nova. Further, novae are known to evolve with 
 time towards higher 
 temperatures which characterize the nebular and coronal phases. Such  
 behavior is not seen here. It also does not appear to be a recurrent nova
  because
 no earlier outbursts have been recorded or reported for 
  V445 Puppis. In a 
 symbiotic nova,  one expects to see
 high-excitation emission lines in the spectrum during the decline
  phase - a feature which
 is absent  here (Kenyon 1986). Also, the light curves of symbiotic novae 
 decline 
 much slower than what is observed in V445 Puppis. Most importantly,
  hydrogen-deficiency is not expected in recurrent or  symbiotic novae.\\
 
 Another possibility is that  V445 Puppis is a 
 born-again AGB star. The nuclei of intermediate mass stars, evolving from 
 the AGB   stage into   planetary nebulae  can experience a final 
 Helium shell  flash which lifts them into a high luminosity range again 
  (Iben et al. 1983). There 
  are only 3 known examples of born-again AGB's 
  in the Galaxy viz. V605 Aql, FG Sge and 
  Sakurai's   object. However there are a few arguments against V445 Puppis 
  being a born-again AGB. First, no nebulosity (caused by gas ionized by
 the hot nuclei) has been seen around the object as has been found in the
  other known cases. We have checked one of the H$\alpha$ surveys covering 
  this region (Schwartz 1990) but  the object is not listed as showing
  signs of emission. Again, born-again AGB's - unlike
 V445 Puppis-  brighten  very gradually to their peak brightness - a process 
 which can even take decades. For the actual timescales observed, 
 the reader may refer to Duerbeck et al. (2000) for Sakurai's object,
  Harrison (1996) for V605 Aql and Fig. 1 of  Blocker \& Schonberner (1997)
  for FG Sge.\\ 
 
  There is a class of eruptive variables whose spectrum after the outburst 
 evolves into that of a cool M-type giant or supergiant. This  spectral 
 evolution
 takes place rather rapidly - within a time  of 1 to 2 months.
  The first such object
 in this class is  a  luminous, red variable star
   (M31 RV) that erupted in M31 (Rich 
 et al. 1989;  Mould et al. 1990). Subsequently the eruptive variable V4332 Sgr
  (Martini et al. 1999) and most recently V838 Mon, (Banerjee \& Ashok 2002a;
   Munari et al. 2002) have been added to this
class. A consensus has not been reached on placing these objects in the same
category but what is certain is that after a nova-like outburst they quickly
 evolve to a   very cool M type (or even later type) absorption spectrum. 
 V838 Mon has, in fact, evolved at present to be similar to that of  a
 T type brown dwarf 
 ( Banerjee \& Ashok 2002b, Geballe et al. 2002).  In the case of V445 Puppis,
 there is VSNET photometric data in the $BVRI$ bands for
 more than 5 months after it's outburst. We have checked these colors
  and do not find an evolutionary trend towards M type
  giant/supergiant stage. For e.g. on 2001, March 30.48  ( more than 4 months
 after the outburst) the observed magnitudes are
 $B$ = 10.66 and $V$ = 10.13. For an adopted value of E(B-V) = 0.25,
 this results into an intrinsic (B-V) color of 0.29 which is very different
 from $\sim$ 1.6 which is the expected value for a cool M giant/supergiant 
 star. Thus V445 Puppis
 does not appear to show an  evolutionary trend 
 similar to the V838 Mon class of objects.\\
 
 V445 Puppis also does not appear to be an RCB or Hydrogen deficient Carbon 
 star (HdC) . RCB stars  show episodic dimming due to ejection of dusty
  carbon shells, subsequently followed by a rebrightening. But their spectra 
 are typically F type absorption spectra and
 most of them are detected in IRAS and show an IR excess. V445 Puppis was
 not detected previously by IRAS. Furthermore, no other outbursts  
  have been recorded for V445 Puppis. HdC stars share most of the 
 properties of RCB stars except that they  do not show large
  brightness 
 variations. Both groups of objects have absolute magnitudes $M{_{\rm V}}$
 in the range -3 to -5 (Brunner et al. 1998). Unless the extinction
   $A{_{\rm v}}$ and distance 
 $d$ to V445 Puppis are grossly underestimated, $M{_{\rm V}}$  $\sim$ 0 for 
 V445 Puppis. This is inferred from 
 $m{_{\rm v}}$ - $M{_{\rm V}}$ = 5log$d$ - 5 + $A{_{\rm v}}$
  using adopted values of $A{_{\rm v}}$ $\sim$ 1,
  an upper limit for $d$ = 3 kpc and the pre-outburst magnitude value of 
  $m{_{\rm v}}$ = 13.6. The object appears  under luminous for an RCB star.\\

 The possibility that V445 Puppis is not a single star - but part of a binary
 system with an accretion disk - has some supporting  evidence. First,
  the SED  of the star in the pre-outburst stage (Fig. 7, lower panel)
   is difficult to fit
 with a   black body spectrum. We have tried several black body fits
at different temperatures to the observed SED. While fitting, the observed 
SED has itself been modified by choosing different $E(B-V)$ values in case the
 adopted value of $E(B-V)$ = 0.25 is grossly incorrect. However a good fit 
 is not
found and the closest fit to the data ( $T$ = 6500K, $E(B-V)$ = 0.25)
  is
 not satisfactory as can be seen from Fig. 7 (lower panel). It is
   possible that 
 the observed continuum distribution from V445 Puppis is dominated by radiation
  from an accretion disk surrounding the object. In the case of a steady-state
  accretion disk around a white-dwarf, the continuum 
  radiation from the disk can be described by an
   $F{_{\rm \nu}}$ $\alpha$ $\nu{^{\rm 1/3}}$ relation 
 ( Mayo et al. 1980 and other references therein).
  A $F{_{\rm \nu}}$ $\alpha$ $\nu{^{\rm 1/3}}$  dependence
  falls off less steeply than the $F{_{\rm \nu}}$ $\alpha$ $\nu{^{\rm 2}}$ 
  relation expected for a black body in the Rayleigh-Jeans regime. We find 
  that such an 
 accretion disk spectrum fits the data of Fig. 7 much better than a black body
 fit.\\
 
  The radio data from Rupen et al. (2001a, b and their website 
 http://www.aoc.nrao.edu/~mrupen/XRT/V445Pup ) also 
 suggest the possibility of an  accretion process in a binary system. 
  They state that  the
observed synchrotron emission  probably originates
   either 
 in accretion onto a compact companion  or in an asymmetric, clumpy shock at 
 the boundary between ejecta  from a fresh outburst in September 2001 and
  previous ejecta (the  fresh  outburst would be  obscured by the thick dust
  shell which had formed around  this time). But they argue that the rapid 
 evolution of both the radio   emission and the radio absorption - significant
 flaring is seen in the object - favors the presence of an accretion process.
   However, it must be pointed out,  there is a lacuna in   comparing the
observed radio emission with radiation from an accretion disk with 
a $\nu{^{\rm 1/3}}$ spectra as discussed above. The latter does not constitute a
non-thermal  source like the  observed radio emission. But the present
radio detection was made well after the outburst. It may just be possible
that the physical conditions/mechanisms of the accretion process are different
in the pre- and post-outburst stages leading to a qualitative difference in
the radiation emitted. We are unable to judge on this aspect. The primary
 aim of the present work was to look for any evidence - albeit even 
 suggestive - 
which would show whether V445 Puppis belongs to a binary system or not. Given 
the strangeness of the object, this would help  understand it's nature. We feel
the recent radio data and the pre-outburst SED seem to indicate a
 binary nature for V445 Puppis.\\

  From the arguments given above, we find that it is
  difficult to place V445 Puppis into known categories of eruptive 
 variables. Kato et al. (1989) and Iben \& Tutukov (1994), among other 
 workers,  have 
 investigated the case of Helium novae. Kato et al. (1989)
 consider the
 case of a degenerate white dwarf accreting Helium from its 
 Helium rich 
 companion. For
 appropriate mass accretion rates, an accretion-induced collapse of the white
 dwarf into a neutron star need not occur. Instead a thermo-nuclear reaction 
 can occur on the surface of the white dwarf leading to a Helium nova outburst.
 The ejecta of such an outburst, in the Kato et al. (1989) model, should
  presumably be rich  in Helium and Carbon and highly 
  hydrogen-deficient. This is based  on the
 assumed, pre-ignition, chemical composition of the envelope by mass viz 
 $X$ = 0, $Y$ = 0.97 and $Z$ = 0.03. Further, Carbon in the ejecta is 
 expected to be additionally enhanced because the thermonuclear outburst
  process itself converts  Helium to Carbon. Such a scenario, or some close
  variant of it,  may apply to V445 Puppis and explain the hydrogen-deficiency 
 ( and He/C enrichment) of it optical and IR spectra. The tentative evidence
  for binarity and  accretion disk in V445 Puppis is also in conformity 
  with the requirements of 
 a helium nova scenario. However, the predicted luminosity of the outburst
 in a Helium nova is $L$ $\sim$ $10{^{\rm 5}}$$L$$_\odot$ (Kato et al. 1989)
  and X-rays are also predicted in
 the post-outburst stage (Iben \& Tutukov 1994). In the case of V445 Puppis,
 there are no 
 reports on  X-ray detection from the object, though it is uncertain whether
  any attempt has been made in this direction.   Further the thick dust
   shell  may attenuate any soft X-ray emission that takes place. The
    outburst luminosity, if
   approximated from the SED of 2 Jan. 01 (Fig. 7), is found to be closer to
 $10{^{\rm 4}}$$L$$_\odot$ i.e. slightly less than the theoretical estimate.
 In view of this,  it may be slightly speculative to identify V445 Puppis 
 as a Helium nova, but similarities exist. Given the interesting and rather
 unique properties of the object, it is hoped that this work will prompt a 
 rigorous, theoretical study to explain the nature of V445 Puppis.

\begin{acknowledgements}
	  
	  The research work at Physical Research Laboratory is funded by
the Department of Space, Government of India. We thank A.Tej for help in
obtaining observations.  We thank R.M. Wagner, the referee, for useful
and constructive comments that have helped improve the paper.
This work has made use of
 data available from 2MASS data center and also from data/information 
 available at the following websites viz. 
 \protect http://kurucz.harvard.edu/linelists.html; 
 \protect http://vela.as.arizona.edu/$\sim$rmw/v445pup.html;
 \protect http://www.kusastro.kyoto-u.ac.jp.vsnet,
 \protect http://www.aoc.nrao.edu/$\sim$mrupen/XRT/V445Pup and
 \protect http://www1.harenet.ne.jp/~aikow/0113pnp.gif 
\end{acknowledgements}


\end{document}